\documentclass[twocolumn,floatfix, superscriptaddress,showpacs,]{revtex4}
\usepackage{graphicx}
\usepackage{amsmath}
\usepackage{bm}
\begin{document}

\title{Valley filter and valley valve in graphene}
\author{A. Rycerz}
\affiliation{Instituut-Lorentz, Universiteit Leiden, P.O. Box 9506, 2300 RA Leiden, The Netherlands}
\affiliation{Marian Smoluchowski Institute of Physics, Jagiellonian University, Reymonta 4, 30--059 Krak\'{o}w, Poland}
\author{J. Tworzyd{\l}o}
\affiliation{Institute of Theoretical Physics, Warsaw University, Ho\.{z}a 69, 00--681 Warsaw, Poland}
\author{C. W. J. Beenakker}
\affiliation{Instituut-Lorentz, Universiteit Leiden, P.O. Box 9506, 2300 RA Leiden, The Netherlands}
\date{August 2006}
\begin{abstract}
It is known that the lowest propagating mode in a narrow ballistic ribbon of graphene may lack the twofold valley degeneracy of higher modes. Depending on the crystallographic orientation of the ribbon axis, the lowest mode mixes both valleys or lies predominantly in a single valley (chosen by the direction of propagation). We show, using a tight-binding model calculation, that a nonequilibrium valley polarization can be realized in a sheet of graphene, upon injection of current through a ballistic point contact with zigzag edges. The polarity can be inverted by local application of a gate voltage to the point contact region. Two valley filters in series may function as an electrostatically controlled ``valley valve'', representing a zero-magnetic-field counterpart to the familiar spin valve.
\end{abstract}
\maketitle

The potential of graphene for carbon electronics rests on the possibilities offered by its unusual band structure to create devices that have no analogue in silicon-based electronics \cite{Nov04,Ber04}. Conduction and valence bands in graphene form conically shaped valleys, touching at a point called the Dirac point. There are two inequivalent Dirac points in the Brillouin zone, related by time-reversal symmetry. Electrons and holes in each valley are massless, with a large energy-independent velocity $v=10^{6}\,{\rm m}/{\rm s}$. Intervalley scattering is suppressed in pure samples \cite{Gui06,Mor06,McC06}. The independence and degeneracy of the valley degree of freedom suggests that it might be used to control an electronic device, in much the same way as the electron spin is used in spintronics \cite{Wol01} or quantum computing \cite{Cer05}.  

A key ingredient for ``valleytronics'' would be a controllable way of occupying a single valley in graphene, thereby producing a valley polarization. Different lines of approach are possible: One might use the fact that the lowest Landau level in a strong magnetic field is valley polarized \cite{Nov05,Zha05}. In zero magnetic field, the exchange interaction may lead to a spontaneous valley polarization \cite{Gun06}. Here we propose an electrostatic method of valley polarization, that relies on the breaking of time-reversal symmetry by a transport current.

Earlier work \cite{Fuj96,Nak96,Wak01,Wak02,Per06,Bre06,Two06} on graphene ribbons (long and narrow ballistic strips) has shown that they may support a propagating mode arbitrarily close to the Dirac point, and that this mode lacks the valley degeneracy of modes that propagate at higher energies. For armchair edges of the ribbon, this lowest propagating mode is constructed from states in both valleys, but for zigzag edges only a single valley contributes \cite{Fuj96,Nak96,Wak01,Wak02,Per06}. In accord with time-reversal symmmetry, the mode switches from one valley to the other upon changing the direction of propagation.

\begin{figure}[tb]
\centerline{\includegraphics[width=0.9\linewidth]{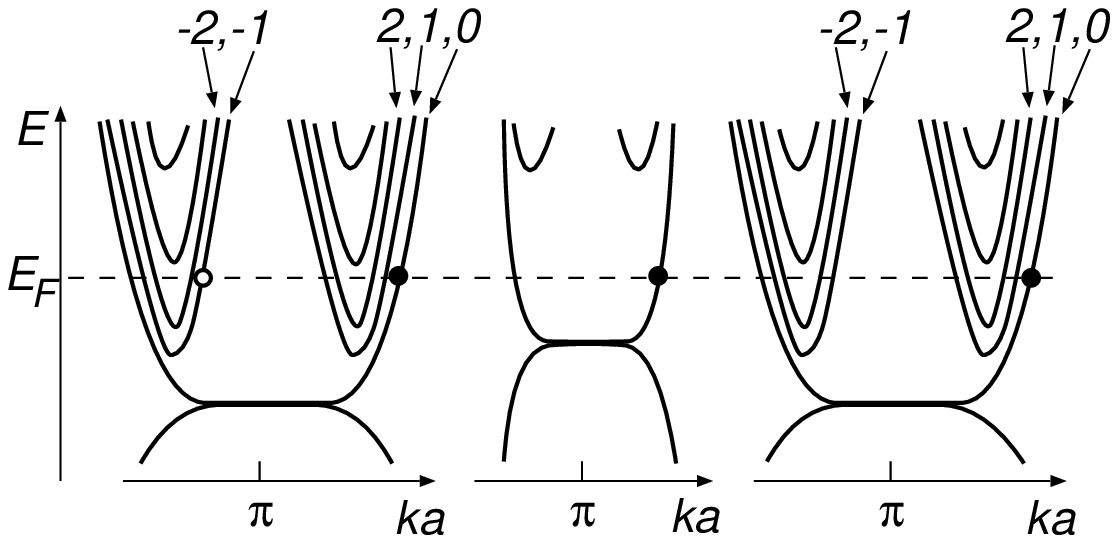}}\medskip

\centerline{\includegraphics[width=0.9\linewidth]{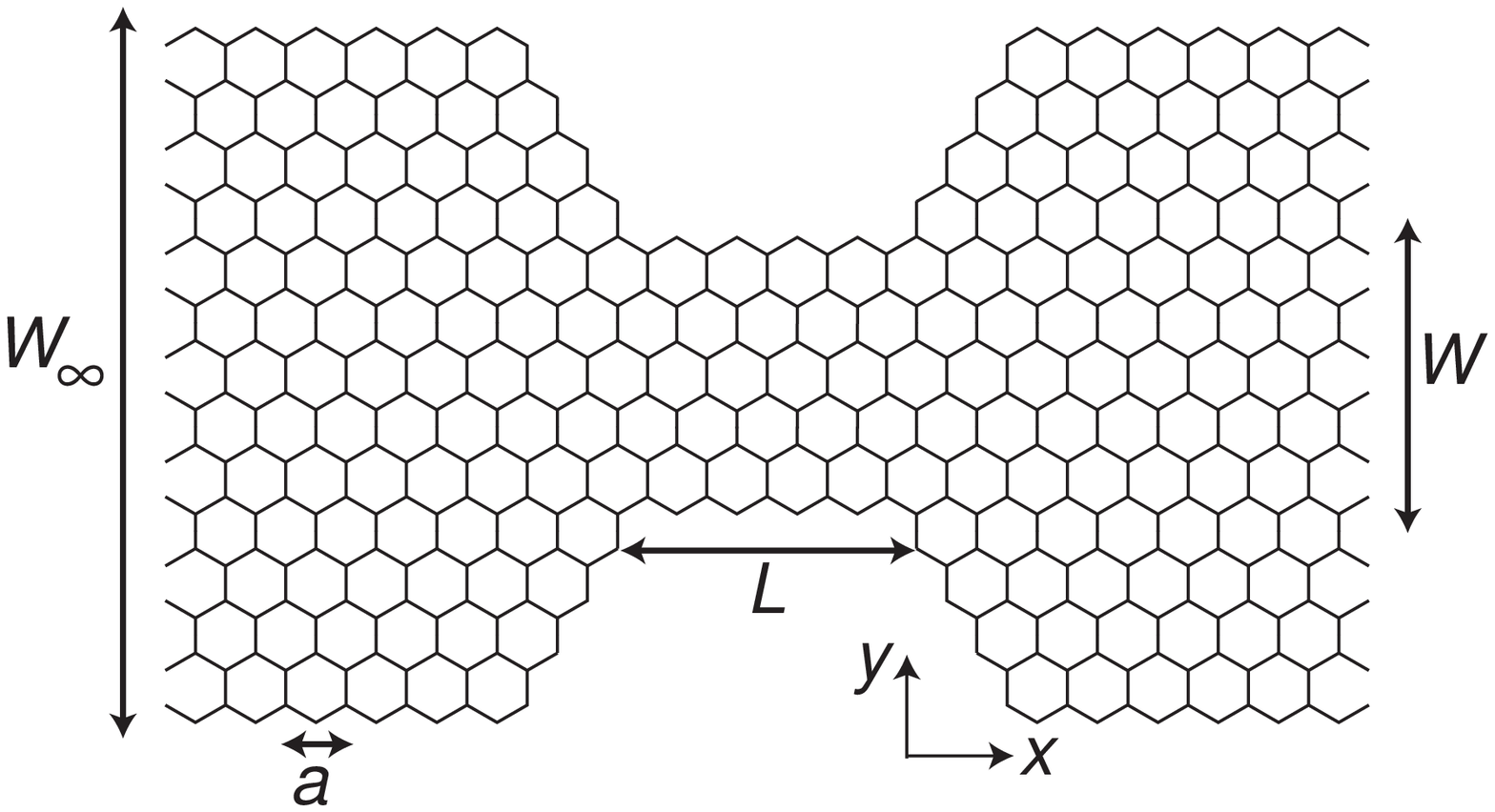}}\medskip

\centerline{\includegraphics[width=0.9\linewidth]{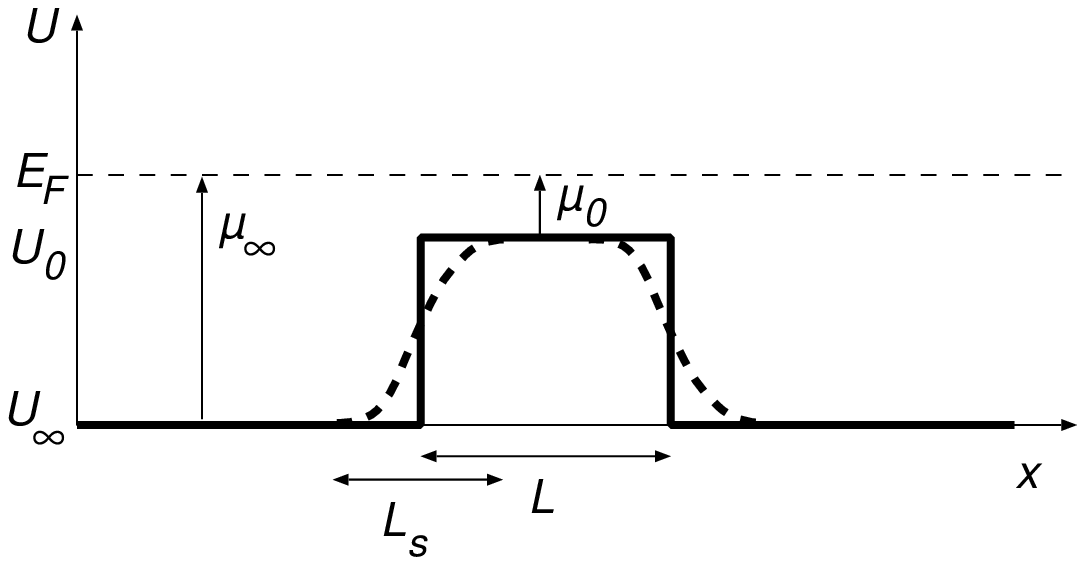}}

\caption{\label{fig_zigzagQPC}
Schematic of the valley filter. Central panel: Honeycomb lattice of carbon atoms in a strip containing a constriction with zigzag edges. Top panel: Dispersion relation in the wide and narrow regions. An electron in the first valley (modes $n=0,1,2,\ldots$) is transmitted (filled circle), while an electron in the second valley (modes $n=-1,-2,\ldots$) is reflected (open circle). Bottom panel: Variation of the electrostatic potential along the strip, for the two cases of an abrupt and smooth potential barrier (solid and dashed lines). The polarity of the valley filter switches when the potential height $U_{0}$ in the constriction crosses the Fermi energy $E_{F}$.}
\end{figure}

In this work we consider a quantum point contact (QPC), which is a short and narrow constriction with a quantized conductance $G=n\times 2e^{2}/h$ \cite{Hou96}. (The factor of two accounts for the spin degeneracy.) A current $I$ is passed through the QPC by application of a voltage difference $V$ between wide regions on opposite sides of the constriction (see Fig.\ \ref{fig_zigzagQPC}). The orientation of the graphene lattice is such that the constriction has zigzag edges along the direction of current flow. We demonstrate by numerical simulation that on the first conductance plateau the QPC produces a strong polarization of the valleys in the wide regions. This finding signifies that the valley polarization present inside the QPC is not destroyed by intervalley scattering at the transition from a narrow to a wide region.

\begin{figure}[tb]
\centerline{\includegraphics[width=0.9\linewidth]{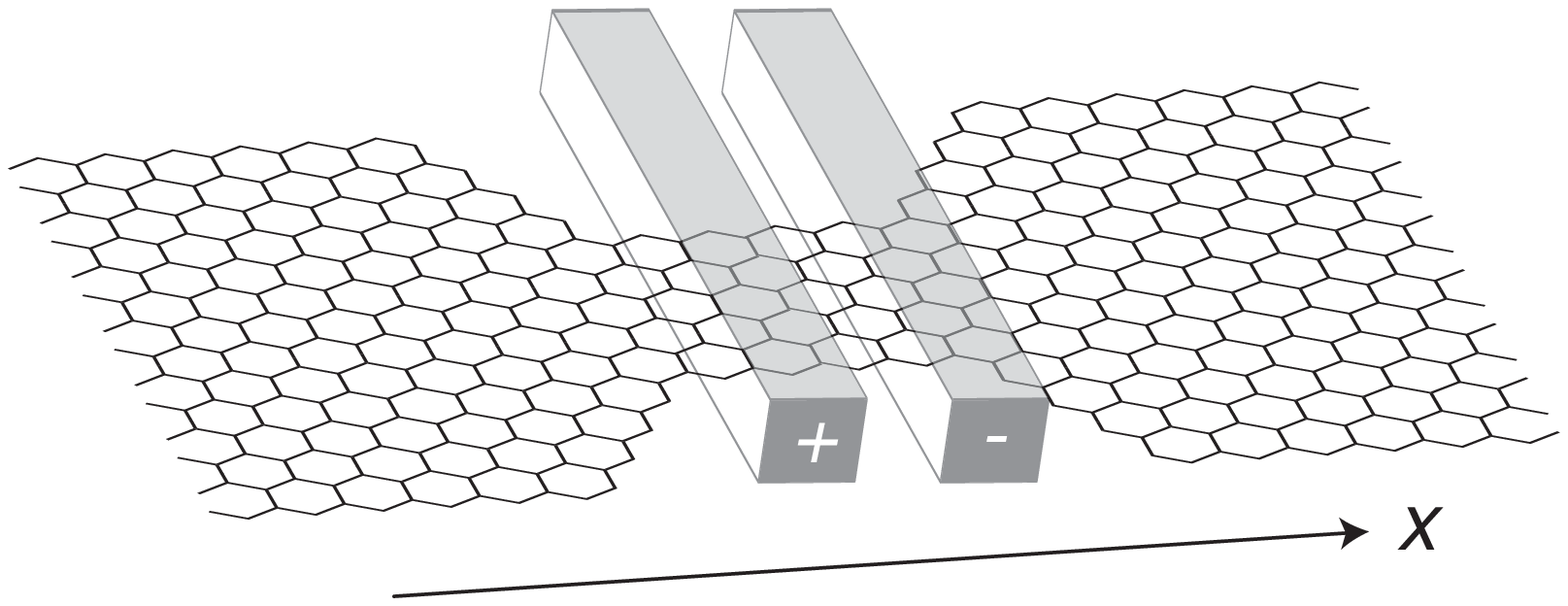}}

\centerline{\includegraphics[width=0.9\linewidth]{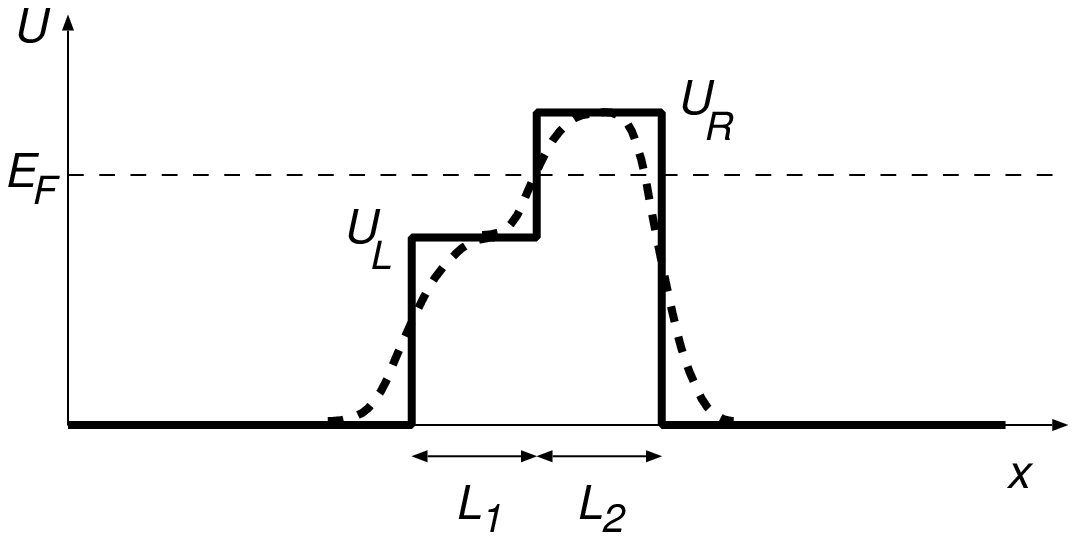}}

\caption{\label{fig_zigzagQPC_3D_2}
Schematic of the valley valve (top) and corresponding potential profile (bottom). The current through the constriction is blocked if $\mu_{L}=E_{F}-U_{L}$ and $\mu_{R}=E_{F}-U_{R}$ have opposite sign.}
\end{figure}

We show that the polarization of this valley filter can be inverted by locally raising the Dirac point in the region of the constriction, by means of a gate voltage, such that the Fermi level lies in the conduction bands in the wide regions and in the valence band inside the constriction. Two valley filters in series, one acting as a polarizer and the other as an analyzer, can block the current if they have the opposite polarity (see Fig.\ \ref{fig_zigzagQPC_3D_2}). This extends earlier findings on the effect of a potential barrier in a graphene ribbon by Wakabayashi and Aoki \cite{Wak02},  demonstrating that a QPC can operate as a ``valley valve'' --- a purely electronic counterpart of the magneto-electronic spin valve.

Our calculations start from the tight-binding model of graphene, with Hamiltonian
\begin{equation}\label{Hdef}
H=\sum_{i,j}\tau_{ij}|i\rangle\langle j|+\sum_{i}U_{i}|i\rangle\langle i|.
\end{equation}
The hopping matrix element $\tau_{ij}=-\tau$ if the orbitals $|i\rangle$ and $|j\rangle$ are nearest neigbors on the honeycomb lattice, otherwise $\tau_{ij}=0$. The electrostatic potential energy $U_{i}=U(x_{i})$ varies only along the axis of the constriction. It equals $U_{\infty}$ in the wide regions and rises to $U_{0}$ inside the constriction. We smooth the step wise increase of the potential over a length $L_{s}$, according to the function
\begin{equation}
\Theta_{L_{s}}(x)=\left\{\begin{array}{cl}
0&{\rm if}\;\;x<-L_{s}/2,\\
\tfrac{1}{2}+\tfrac{1}{2}\sin(\pi x/L_{s})&{\rm if}\;\;|x|<L_{s}/2,\\
1&{\rm if}\;\;x>L_{s}/2.
\end{array}\right.\label{thetadef}
\end{equation}
The potential barrier $U(x)=U_{\infty}+(U_{0}-U_{\infty})[\Theta_{L_{s}}(x)-\Theta_{L_{s}}(x-L)]$ is rectangular for $L_{s}=0$ (solid line in Fig.\ \ref{fig_zigzagQPC}, bottom panel), while it has a sinus shape for $L_{s}=L$ (dashed line).

\begin{figure}[tb]
\centerline{\includegraphics[width=0.9\linewidth]{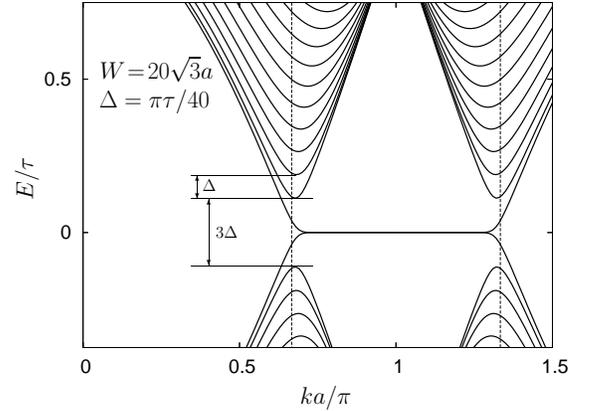}}
\caption{\label{fig_zband}
Dispersion relation of a graphene strip with zigzag edges. The spacing of the low-lying modes approaches $\Delta\equiv\frac{1}{2}\sqrt{3}\,\pi\tau a/W$ for $W/a\gg 1$. The zeroth and first modes have a larger spacing, approaching $3\Delta/2$ for $W/a\gg 1$. Vertical lines mark the valley centers at $k=2\pi/3a$ and 
$4\pi/3a$.
}
\end{figure}

The dispersion relation of the honeycomb lattice in a strip with zigzag edges is shown schematically in Fig.\ \ref{fig_zigzagQPC} (top panel) and exactly in Fig.\ \ref{fig_zband}. The wide regions support $2N+1$ propagating modes at the Fermi energy $E_{F}$, which form a basis for the transmission matrix $t$. Modes $n=1,2,\ldots N$ lie in the first valley [with longitudinal wave vector $ka\in(\pi,2\pi)$] while modes $n=-1,-2,\ldots -N$ lie in the second valley [with $ka\in(0,\pi)$]. The zeroth mode $n=0$ lies in a single valley fixed by the direction of propagation. The conductance of the constriction is determined by the Landauer formula
\begin{equation}
G=\frac{2e^{2}}{h}\sum_{n=-N}^{N}T_{n},\;\;T_{n}=\sum_{m=-N}^{N}|t_{nm}|^{2}.\label{Landauer}
\end{equation}
The valley polarization of the transmitted current is quantified by
\begin{equation}
P=\frac{T_{0}+\sum_{n=1}^{N}(T_{n}-T_{-n})}{\sum_{n=-N}^{N}T_{n}},\label{Pdef}
\end{equation}
where we consider the case (illustrated in Fig.\ \ref{fig_zigzagQPC}) that the zeroth mode lies in the first valley. The polarization $P\in[-1,1]$, with $P=1$ if the transmitted current lies fully in the first valley and $P=-1$ if it lies fully in the second valley.

\begin{figure}[tb]
\centerline{\includegraphics[width=0.9\linewidth]{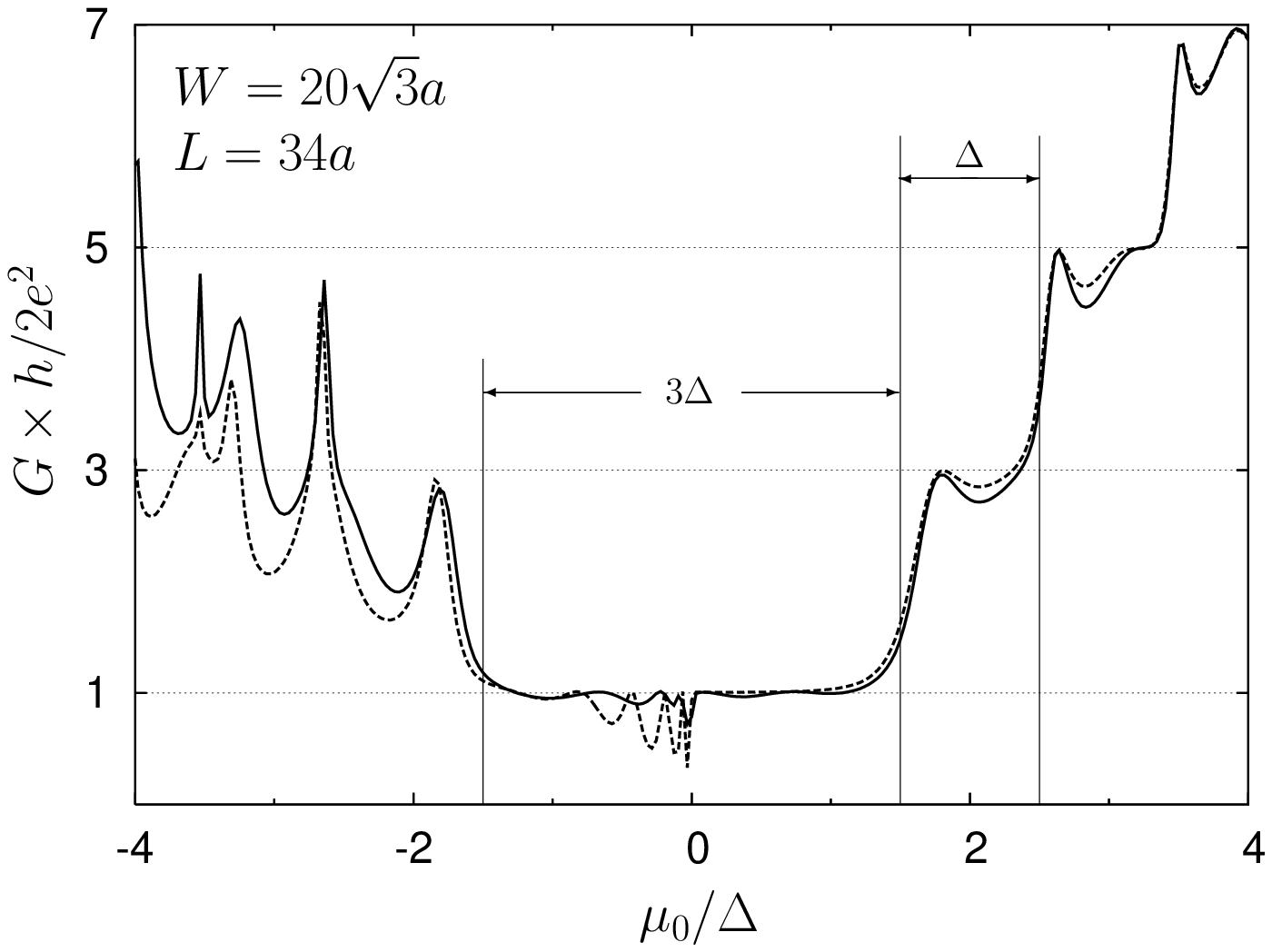}}

\centerline{\includegraphics[width=0.9\linewidth]{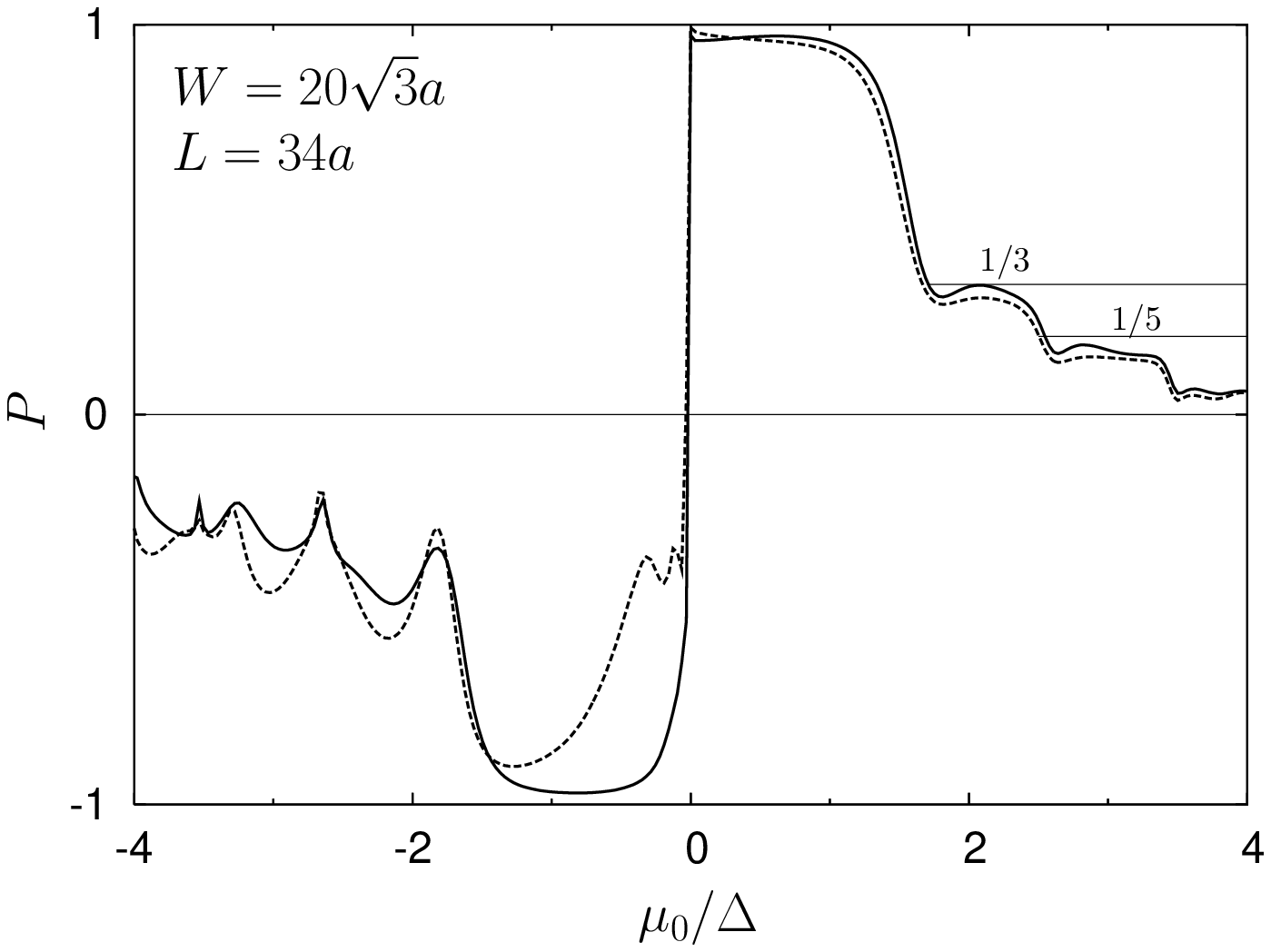}}
\caption{\label{fig_gquant}
Conductance (top panel) and valley polarization (bottom panel) for the valley filter of Fig.\ \ref{fig_zigzagQPC}, as a function of the electrochemical potential in the narrow region. The solid and dashed lines corresponds to abrupt ($L_s=0$) and smooth ($L_s=8a$) potential steps, respectively.
}
\end{figure}

\begin{figure}[tb]
\centerline{\includegraphics[width=0.9\linewidth]{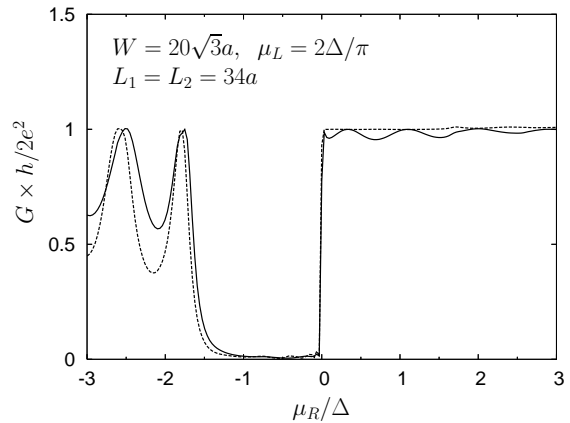}}
\caption{\label{fig_gvalve}
Conductance for the valley valve of Fig.\ \ref{fig_zigzagQPC_3D_2} at fixed $\mu_{L}$ as a function of $\mu_{R}$. The solid and dashed lines correspond to abrupt ($L_s=0$) and smooth ($L_s=8a$) potential steps, respectively.
}
\end{figure}

We have calculated the transmission matrix numerically by adapting to the honeycomb lattice the method developed by Ando for a square lattice \cite{And91}. Results are shown in Figs.\ \ref{fig_gquant} and \ref{fig_gvalve}. We have fixed the width of the wide regions at $W_{\infty}=70\sqrt{3}\, a$ (in units of the lattice spacing $a$). The electrochemical potential in the wide regions is set at $E_{F}-U_{\infty}\equiv\mu_{\infty}=\tau/3$, corresponding to $2N+1=29$ propagating modes. The narrow region has width $W=20\sqrt{3}\,a$. We measure the electrochemical potential $E_{F}-U_{0}\equiv\mu_{0}$ in the narrow region in units of the mode spacing $\Delta\equiv\frac{1}{2}\sqrt{3}\,\pi\tau a/W=\pi\tau/40$ (indicated in Fig.\ \ref{fig_zband}).

The operation of the {\em valley filter\/} is demonstrated in Fig.\ \ref{fig_gquant}. The top panel shows the conductance, while the bottom panel shows the valley polarization --- both as a function of the electrochemical potential $\mu_{0}$ in the narrow region. For positive $\mu_{0}$ the current flows entirely within the conduction band, and we obtain plateaus of quantized conductance at odd multiples of $2e^{2}/h$ (as predicted by Peres et al.\ \cite{Per06}). Smoothing of the potential step improves the flatness of the plateaus (compare solid and dashed lines). The plateaus in the conductance at $G=(2n+1)\times 2e^{2}/h$ correspond to plateaus in the valley polarization at $P=1/(2n+1)$. On the lowest $n=0$ plateau, and for $0<\mu_{0}\lesssim\Delta$, the polarization is more than 95\%. 

For negative $\mu_{0}$ the current makes a transition from the conduction band in the wide regions to the valence band in the narrow region. This interband transition has previously been studied in an unbounded system \cite{Che06,Kat06}, where it leads to selective transmission at normal incidence. In the QPC studied here we find that the interband transition destroys the conductance quantization --- except on the first plateau, which remains quite flat in the entire interval $-3\Delta/2<\mu_{0}<3\Delta/2$. The resonances at negative $\mu_{0}$ are due to quasi-bound states in the valence band \cite{Mil06,Sil06}. The polarity of the valley filter is inverted for negative $\mu_{0}$, with some loss of quality (in particular for the smooth potential).

The operation of the {\em valley valve\/} is demonstrated in Fig.\ \ref{fig_gvalve}. The current is blocked for $-3\Delta/2<\mu_{R}<0$ with $\mu_{L}$ on the first conductance plateau, so that the constriction contains two valley filters of opposite polarity in series. The switching behavior of Fig.\ \ref{fig_gvalve} is similar to that obtained by Wakabayashi and Aoki in a simulation of a zigzag graphene ribbon containing a potential barrier \cite{Wak02}. Their geometry does not have well-separated valleys (since there is no wide multi-mode region), so an interpretation in terms of valley filtering is not as obvious as in the wide-narrow-wide geometry considered here.

In conclusion, we have demonstrated by computer simulation that a 
quantum point contact with zizag edges is capable of producing a highly 
nonequilibrium population of the valleys in a carbon monolayer. The 
valley polarization is better than 95\% if the Fermi level lies above 
the Dirac point both inside and outside of the constriction. The 
polarity of the valley filter is inverted when the Fermi level in the 
wide and narrow regions straddles the Dirac point, permitting the 
construction of a switch (valley valve) consisting of two filters in 
series. We anticipate that the experimental realization of the device 
proposed in this work will make it possible to exploit the valley degree 
of freedom, in addition to spin and charge degrees of freedom, as a 
carrier of information in carbon electronics.

This research was supported by the Dutch Science Foundation NWO/FOM and by the European Community's Marie Curie Research Training Network (contract MRTN-CT-2003-504574, Fundamentals of Nanoelectronics). AR acknowledges a Foreign Postdoc Fellowship from the Polish Science Foundation (FNP) and support by the Polish Ministry of Science (Grant No.\ 1~P03B~001~29).

\end{document}